# OTFDM: A Novel 2D Modulation Waveform Modeling Dot-product Doubly-selective Channel

Yihua Ma[1,2], Zhifeng Yuan[1,2], Yu Xin[1,2], Jiang Hua[1,2], Guanghui Yu[1,2], Jin Xu[1,2], Liujun Hu[1,2]
[1]State Key Laboratory of Mobile Network and Mobile Multimedia Technology, [2]ZTE Corporation
Shenzhen, China
{yihua.ma, yuan.zhifeng, xin.yu, hua.jian2, yu.guanghui, xu.jin7, hu.liujun}@zte.com.cn

*Abstract*—Recently, a two-dimension (2D) modulation waveform of orthogonal time-frequency-space (OTFS) has been a popular 6G candidate to replace existing orthogonal frequency division multiplexing (OFDM). The extensive OTFS researches help to make both the advantages and limitations of OTFS more and more clear. The limitations are not easy to overcome as they come from OTFS on-grid 2D convolution channel model. Instead of solving OTFS inborn challenges, this paper proposes a novel 2D modulation waveform named orthogonal time-frequency division multiplexing (OTFDM). OTFDM uses a 2D dot-product channel model to cope with doubly-selectivity. Compared with OTFS, OTFDM supports grid-free channel delay and Doppler and gains a simple and efficient 2D equalization. The concise dot-division equalization can be easily combined with MIMO. The simulation result shows that OTFDM is able to bear high mobility and greatly outperforms OFDM in doubly-selective channel.

## I. INTRODUCTION

Air interference evolution has always been an important aspect of different generations of communication systems. In both 4G and 5G, orthogonal frequency division multiplexing (OFDM) [1] waveform was standardized and provides a great improvement compared with code-division multiplexing access (CDMA) in 3G. Orthogonal time frequency space (OTFS) [2][3] has been very popular in 6G waveform research, which still has many fundamental issues to tackle. This paper proposed a novel two-dimension (2D) modulation waveform, orthogonal time-frequency division multiplexing (OTFDM), to enhance OFDM and avoid the problems of OTFS.

OFDM has many advantages in terms of spectral efficiency, flexibility, and multi-input and multi-output (MIMO) compatibility. One crucial advantage of OFDM is the simple channel equalization, which turns the complex convolutional multi-path channel into the simple dot product. This advantage also helps to realize a simple and efficient MIMO multi-user (MU) transmission. The one-dimensional dot-product channel model fails for the time-selective channel. 5G NR can shorten the OFDM symbol to make Doppler neglectable. However, inter-symbols interference occurs if cyclic prefix (CP) is shortened, while spectral efficiency (SE) degrades if CP becomes longer and the OFDM symbol is shorter. This disadvantage has been revealed as the major drawback of OFDM in a fair comparison to OTFS [4]. That is to say, the SE of OFDM degrades in a high-mobility doubly-selective channel.

OTFS was proposed for 5G discussion in 3GPP 84-bis meeting [5]. Compared with OFDM, OTFS supports a higher mobility and have the so-called channel diversity gain. However, the gain of channel diversity is tiny as channel coding is a must to resist fading. The OTFS orthogonality at the receiver vanishes and cannot be approximately obtained. To solve non-orthogonality data, message-passing algorithms (MPA) [6][7] are usually required. Fortunately, the delay-Doppler (DD) domain non-orthogonal spreading pattern is relatively sparse, which gains a low-complexity MPA. To further reduce complexity, linear solvers have been proposed for OTFS [8] with acceptable performance degradation and on-grid delay-Doppler assumption. However, the complexity is still much higher than the simple FFT operation of the approximately orthogonal OFDM receiver. The complexity limits the data throughput of real-time baseband processing. Moreover, the large OTFS symbol leads to large latency.

The limitation of OTFS comes from the inborn nature of grid-based channel assumption and complex convolutional channel model. As a comparison, OFDM supports grid-less delays and uses a very simple dot-product channel model. To keep these advantages in the doubly-selective channel, this paper proposes OTFDM to support the practical grid-less delay-Doppler and obtain a simple 2D dot-product channel. OTFDM acts as an extra option to the existing OFDM-based scheme. When OTFDM is enabled, a Doppler spreading and Doppler dot-product are used. This paper also proves that the data part of OTFDM is equivalent to OFDM with comb interleaving. The discrete Fourier transform spreading (DFT-s-) OTFDM is similar to OTFS with an additional Doppler dot-product. Thus, OTFDM is a bridge that connects OFDM and OTFS. Morever, the pilot insertion and channel estimation scheme are designed, and the pilot overhead is analyzed. The analysis shows that the pilot overheads of OFDM, OTFS, and OTFDM are similar as they requires the same time-frequency degrees of freedom to cope with the doubly-selective channel. Also, two Doppler despreading schemes at the receiver are proposed for OTFDM. At last, a link-level simulation is done using taped delay line type C (TDL-C) channel model with a large delay spread and high mobility. The simulation results show that the proposed OTFDM performs much better than OFDM with a low-complexity approximately orthogonal receiver.

The remainder of this paper is organized in the following way. Section II introduces the math model and further analysis the problems of OTFS. Section III proposes the OTFDM transceiver design and declares the difference between DFT-s-OFDM and OTFS. Section IV proposes the channel estimation scheme and analyzes the pilot overheads. In this paper, $(\cdot)^*$, $(\cdot)^T$, and $(\cdot)^H$ denote conjugate, transpose, and Hermitian transpose of a matrix or vector. $(\cdot)_N$ represents the modulo-$N$ operation. $A[i, j]$ represents the element in the $(i+1)$-th row and $(j+1)$-th column of matrix $\mathbf{A}$. $a[k]$ represents the $(k+1)$-the element of vector $\mathbf{a} = \text{vec}(\mathbf{A})$.

## II. MATH MODEL AND EXISTING WORKS

### A. Channel Model

A two-dimension transport block of $\mathbf{X} \in \mathbb{C}^{M \times N}$ is assumed. Assuming a bandwidth of $B$ and a time duration of $T$, the sampling time $T_S = T/MN$. If an $MN$-point CP-OFDM is used, the sub-carrier spacing is $\Delta f = B/MN = 1/T$, and the transmitting signal vector is

$$\mathbf{s}_a = \frac{1}{\sqrt{MN}} \mathbf{F}_{MN}^H vec(\mathbf{X}) \in \mathbb{C}^{MN \times 1}, \quad (1)$$

where $\mathbf{F}_{MN}$ and $vec(\cdot)$ represent an $MN$-point DFT matrix and vectorization function. A CP of $N_{CP}$ points is added to cover the maximum effective multiple paths.

If a short CP-OFDM symbol of $M$-point IFFT is used to resist high mobility, the transmitted signal vector is

$$\mathbf{s}_b = vec\left(\frac{1}{\sqrt{M}} \mathbf{F}_M^H \mathbf{X}\right) \in \mathbb{C}^{MN \times 1}, \quad (2)$$

which represents $N$-time $M$-point IFFT operation. In this way, the sub-carrier spacing becomes $\Delta f' = B/M = N/T$, and $N$ CPs of total $MN_{CP}$ points are required to be inserted.

Discrete Zak transform (DZT) based generation [3] is a concise generation method for OTFS. Although symplectic finite Fourier transform (SFFT) based generation seems to be compatible with OFDM scheme, part of SFFT cancels with the IFFT of OFDM, which means the so-called compatibility leads to a waste of complexity. Using DZT-based OTFS generation, the transmit signal vector is

$$\mathbf{s}_c = vec(\mathbf{S}_c) = vec\left(\frac{1}{\sqrt{N}} \mathbf{X} \mathbf{F}_N^H\right) \in \mathbb{C}^{MN \times 1}, \quad (3)$$

It represents $M$-time $N$-point IFFT operation. Similar to $MN$-point OFDM, only one CP is added, and the Doppler sub-carrier spacing is also $\Delta f$, instead of $\Delta f'$.

No matter what kind of waveform is selected, the physical channel is the same. Assuming that the channel has $P$ paths, each path has a coefficient $h_i$, a delay of $\tau_i$, and a Doppler shift of $v_i$, $i = 0, 1, ..., P-1$. The received signal and additive Gaussian noise can be either in the vector form $\mathbf{y}$ and $\mathbf{w} \in \mathbb{C}^{MN \times 1}$ or in the matrix form $\mathbf{Y}$ and $\mathbf{W} \in \mathbb{C}^{M \times N}$. With CP, the received signal vector $\mathbf{y}$ is calculated by

$$y[l'] = \sum_{i=0}^{P-1} h_i s\left[(l' - \tau_i/T_s)_{MN}\right] e^{j2\pi(l' - \tau_i/T_s)v_i T_s} + w[l'],$$
$$l' = 0, 1, 2, ..., MN - 1. \quad (4)$$

The on-grid delay is used here just for simplification, which can be easily extended to off-grid cases via oversampling. The on-grid delay and Doppler are required by OTFS, while they are not required by OFDM and OTFDM.

It is not easy to transfer Equation (4) into the DD domain. Quite a few channel models are used in peer-reviewed OTFS papers [3], [6]-[8]. A widely-used DD channel model [6] is

$$Y[l,k] = \sum_{i=0}^{P-1} h_i S\left[(l - \tau_i/T_S)_M, (k - v_i/\Delta f_b)_N\right] e^{j2\pi \tau_i v_i T_S} + W[l,k],$$
$$l = 0, 1, 2, ..., M-1, k = 0, 1, 2, ..., N-1. \quad (5)$$

In the DD channel model, $l$ and $k$ are the index of delay and Doppler bins. Considering the Doppler phase rotation in the delay domain. A relatively accurate model [6] is

$$Y[l,k] = \sum_{i=0}^{P-1} h_i \alpha_{l,k,i} S\left[(l - \tau_i/T_S)_M, (k - v_i/\Delta f)_N\right] + W[l,k],$$

$$\alpha_{l,k,i} = \begin{cases} e^{j2\pi(l-\tau_i/T_S)_M v_i T_S}, l = \tau_i/T_S \sim M-1 \\ e^{j2\pi((l-\tau_i/T_S)_M v_i T_S - (v_i/\Delta f + (k-v_i/\Delta f)_N)M)}, l = 0 \sim \tau_i/T_S - 1 \end{cases}. \quad (6)$$

This model still only works for the on-grid DD domain. When fractional Doppler is considered, a more accurate model can be found in (24) and (25) of [7], which shows the channel gain is actually not constant. Therefore, OTFS does not gain a simple model with practical factors. The approximation seems to be unavoidable to achieve conciseness and efficiency for different waveform schemes.

### B. Inborn Features of OTFS

The challenges of OTFS have already been well summarized in [3]. This paper aims to analyze the root causes of these problems in this sub-section. OTFS does solve the problems of OFDM and only requires one CP for a large symbol, but it was not adopted in the early 5G discussion. Two root causes greatly limit OTFS are the impractical on-grid simplification and the complex convolution channel model.

As shown in Equation (6), Doppler affects the phase rotation of different delay bins. The major power of one off-grid Doppler can be assumed to focus on limited integer Doppler grids, while the phase rotation is still decided by the ground-truth off-grid Doppler frequency. Moreover, it results in non-constant OTFS channel gain as the delay bins at the same Doppler grid could come from different off-grid Doppler frequencies. To alleviate this impact, OTFS requires a large Doppler dimension $N$ to reduce the phase rotation differences between ground-truth off-grid Doppler and neighboring on-grid Doppler. Also, the number of equivalent on-grid paths are hard to decide. A large path number loses the sparsity which greatly increases the complexity. A small path number causes partial energy and unknown interference. Moreover, $M$ cannot be arbitrarily small. A path delay beyond the maximum delay bins is indistinguishable, which means $M \geq \max(\tau_i/T_S)$. Therefore, the block size of OTFS is inevitably large, which leads to large buffering overhead and high transmission latency.

OTFS uses complex deconvolutional channels while OFDM gains simple dot-product channels. Although OTFS uses the sparsity of the DD domain to achieve efficient deconvolution, the complexity is still very high especially for MIMO transceiver still becomes much higher. Before OFDM, the archaic CDMA system deconvolves sparse channels only in the delay domain. CDMA had been evolved to OFDM to get rid of deconvolution, and it seems to be hard to evolve from OFDM to OTFS, which requires a much higher complexity and limited advantages.

## III. THE PROPOSED OTFDM

### A. OTFDM Transmitter

This paper proposes a novel OTFDM scheme that evolves existing OFDM to deal with the complex doubly-

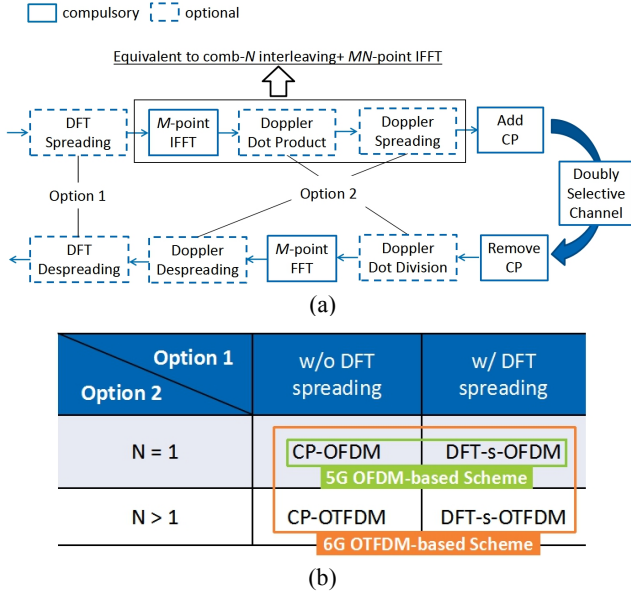

Fig. 1 (a) The transceiver block scheme and (b) option table of OTFDM.

selective channel. Equation (2) has shown OFDM deals with high mobility via using smaller symbols. The problem is that such a method needs $N$ CPs, which greatly reduces SE especially when the CPs are long to cover large delay spread. As a comparison, OTFDM copes with the doubly-selective channel and requires only one CP.

OTFDM is a more generalized OFDM scheme as shown in Fig. 1. If Option 1 is enabled, a DFT spreading is used to get the frequency-domain data $\mathbf{X}$, which does not affect the following processing steps. When $N = 1$, the module becomes the classical OFDM-based scheme. When $N > 1$, novel CP-OTFDM and DFT-s-OTFDM are supported. Option 2 can be supported by setting the values of $N$. The following descriptions are for arbitrary $N$. With the frequency-domain data $\mathbf{X}$, an $M$-point IFFT is employed

$$S_1[l,n] = \frac{1}{\sqrt{M}} \sum_{m=0}^{M-1} X[m,n] e^{j2\pi \frac{ml}{M}}, l = 0,1,...,M-1. \quad (7)$$

Then, a Doppler dot product is used as

$$S_2[l,n] = S_1[l,n] e^{j2\pi \frac{nl}{MN}}. \quad (8)$$

After the dot product, the Doppler spreading is used as

$$S_3[l,k] = \frac{1}{\sqrt{N}} \sum_{n=0}^{N-1} S_2[l,n] e^{j2\pi \frac{nk}{N}}, k = 0,1,...N-1. \quad (9)$$

This step can also be viewed as a transform from Doppler domain to time-domain as Equation (3).

When $N = 1$, Equations (8) and (9) are ineffective, which is transparent for OFDM. Moreover, Equations (7)~(9) are equivalent to a comb-$N$ interleaving and a large $MN$-point IFFT, which represents an $MN$-point large OFDM symbol. The equivalence is proved in Appendix. At last, the transmitter adds one CP and transmits the signal.

*B. OTFDM Receiver*

After removing CP at the receiver side, the receiver obtains the time-domain signal $\mathbf{Y} \in \mathbb{C}^{M \times N}$, or $\mathbf{y} = \text{vec}(\mathbf{Y}) \in \mathbb{C}^{MN \times 1}$, according to the channel model of Equation (4). Note that The notations $n$, $k = 0, 1, 2, ..., N-1$ in OTFDM have different meanings. $n$ is used to represent the $n$-th sub-symbol, e.g., the $n$-th row of $\mathbf{X}$, $\mathbf{S}_1$, or $\mathbf{S}_2$. $k$ is used to represent the $k$-th Doppler spreading replica in OTFDM, e.g., the $k$-th row of $\mathbf{S}_3$. The strategy of OTFDM is to make $M$ small enough to omit the phase rotation caused by Doppler shift within the $k$-th $M$-point Doppler spreading replica.

The contribution of $X[m,n]$ in the received time-domain signal $\mathbf{y}$ is $\gamma_{m,n} \in \mathbb{C}^{MN \times 1}$, which may not locate in the same position at the receiver side. $\gamma_{m,n}$ is used to calculated the spreading coefficients, which will be estimated via pilots in practice. The element of $\gamma_{m,n}$ is

$$\gamma_{m,n}[l'] = \frac{1}{\sqrt{MN}} \sum_{i=0}^{P-1} h_i e^{j2\pi \left( m\Delta f' + \frac{n}{N} \Delta f' + \upsilon_i \right)(l'T_S - \tau_i)} X[m,n],$$
$$l' = 0,1,2,...,MN-1. \quad (10)$$

Then, $\gamma_{m,n}$ is divided into $N$ segments, and $M$ has already been designed to be small enough to make $\upsilon_i$ negligible in $MT_S$ for any $i$. As $\Delta f' = 1/MT_S$, Equation (10) is approximated using $\upsilon_i MT_S \ll 1$ to

$$\gamma_{m,n}[l'] \approx \gamma'_{m,n}[l']$$
$$= \frac{1}{\sqrt{MN}} \sum_{i=0}^{P-1} h_i e^{j2\pi \left[ \frac{(mN+n)(l'T_S - \tau_i)}{MNT_S} + \left\lfloor \frac{l'}{M} \right\rfloor \upsilon_i MT_S - \upsilon_i \tau_i \right]} X[m,n] \quad (11)$$

The Doppler despreading requires to be done at each sub-carrier and each sub-symbol due to doubly selectivity, while the Doppler spreading at the transmitter is done for all sub-carriers of each sub-symbol. Therefore, the receiver-side Doppler despreading is moved to locate after the $M$-point FFT. First, the Doppler dot division is done, which expands $\mathbf{Y}$ to $\mathbf{Y}' \in \mathbb{C}^{M \times N \times N}$ as

$$Y'[l,n,k] = Y[l,k]/e^{j2\pi \frac{nl}{N}}. \quad (12)$$

In this equation, the inter-sub-symbol interference occurs and will be solved by Doppler despreading. With the same dot division, the contribution of transmitted symbol in Equation (11) becomes

$$\gamma'_{m,n}[l']$$
$$= \frac{1}{\sqrt{MN}} \sum_{i=0}^{P-1} h_i e^{j2\pi \left[ \frac{ml'}{M} - \frac{mN+n}{MNT_S}\tau_i + \left\lfloor \frac{l'}{M} \right\rfloor \left( \frac{n}{N} + \upsilon_i MT_S \right) - \upsilon_i \tau_i \right]} X[m,n]. \quad (13)$$

$\mathbf{Y}''$ is obtained from $\mathbf{Y}'$ with an FFT in the first dimension. The element of $\mathbf{Y}''$ is $Y''[m,n,k]$. The $M$-point FFT also combines the time-domain $MN$ points into $N$ points at the sub-carrier $m\Delta f'$ as different sub-carrier using different vectors of $\mathbf{F}_M$. Then, $\gamma''_{m,n} \in \mathbb{C}^{N \times 1}$ whose element is

$$\gamma''_{m,n}[k] = \frac{1}{\sqrt{N}} \sum_{i=0}^{P-1} h_i e^{j2\pi \left[ -\frac{mN+n}{MNT_S}\tau_i + \frac{kn}{N} + \upsilon_i (kMT_S - \tau_i) \right]} X[m,n]. \quad (14)$$

Till this step, $\gamma_{m,n}''[k]$ and $Y''[m,n,k]$ carries the same data information with different index order. The difference is that the former is derived from the transmitted symbol at $m$-th sub-carrier and $n$-th sub-symbol, which is not exactly locating at this position at the receiver side, while $\mathbf{Y}''$ is the received signal that has inter-sub-symbol interference.

Now, a 2D dot-product channel vector $\mathbf{H}_n \in \mathbb{C}^{M \times N}$ for the $n$-th sub-symbol can be obtained as

$$H_n[m,k] = \frac{1}{\sqrt{N}} \sum_{i=0}^{P-1} h_i e^{j2\pi \left[ -\left( \frac{mN+n}{MNT_S} + \upsilon_i \right) \tau_i + \frac{kn}{N} + \upsilon_i kMT_S \right]}. \quad (15)$$

For the $n$-th sub-symbol, $m$ causes time-selective fading, while $l$ causes frequency-selective fading. The 2D dot-product channel provides a new way to deal with the high-mobility doubly-selective channel, which is simpler than OTFS and more robust than OFDM. From Equation (14), the received spreading coefficient at the $m$-th sub-carrier of the $n$-th sub-symbol $\mathbf{c}_{m,n} \in \mathbb{C}^{N \times 1}$ can be expressed by

$$c_{m,n}[k] = \frac{1}{\sqrt{N}} \left( \sum_{i=0}^{P-1} h_i e^{-j2\pi \left( \frac{mN+n}{MNT_S} + \upsilon_i \right) \tau_i} e^{j2\pi \upsilon_i kMT_S} \right) e^{j2\pi \frac{kn}{N}}$$

$$= \frac{1}{\sqrt{N}} \eta_{m,n,k} e^{j2\pi \frac{kn}{N}}. \quad (16)$$

With the time selectivity, the received Doppler spreading codes $\mathbf{c}_{m,n}$ for different sub-symbols are non-orthogonal. In OTFDM, there are two methods to deal with it. The first is time-domain equalization. Each Doppler spreading replica of sub-symbol is divided by the fading coefficient as

$$\gamma'''_{m,n}[k] = \frac{\gamma''_{m,n}[k]}{\eta_{m,n,k}} = \frac{1}{\sqrt{N}} e^{j2\pi \frac{kn}{N}} X[m,n]. \quad (17)$$

After this time-domain equalization, the spreading codes of different sub-symbol at the sub-carrier $\Delta f'$ become orthogonal. The inter-sub-symbol interference in $\mathbf{Y}''$ can be easily distinguished via an FFT as

$$\hat{X}[m,n] = \frac{1}{\sqrt{N}} \sum_{k=0}^{N-1} \frac{1}{\sqrt{N}} e^{-j2\pi \frac{kn}{N}} \frac{Y''[m,n,k]}{\eta_{m,n,k}}. \quad (18)$$

The second method is LMMSE. Equation (15) can be viewed as a MIMO channel. The $N$ Doppler spreading replicas of each sub-symbol can be viewed as $N$ antennas, and the $N$ sub-symbol can be viewed as $N$ users. $M$ sub-carriers are still $M$ sub-carriers. Then, (15) can be used to form a frequency-selective channel $\mathbf{H}_m \in \mathbb{C}^{N \times N}$ at the sub-carrier $m\Delta f'$ as

$$\mathbf{H}_m = \begin{bmatrix} c_{m,1}[1] & c_{m,2}[1] & \dots & c_{m,N}[1] \\ c_{m,1}[2] & c_{m,2}[2] & \dots & c_{m,N}[2] \\ \dots & \dots & \dots & \dots \\ c_{m,1}[N] & c_{m,2}[N] & \dots & c_{m,N}[N] \end{bmatrix}. \quad (19)$$

Assuming the sum of $\mathbf{y}_{m,n}''$ of all $N$ sub-symbols is $\mathbf{y}_m'' \in \mathbb{C}^{N \times 1}$. LMMSE combining $N$ sub-symbols is

$$\hat{X}[m,n] = H_m[m,:] \left( \mathbf{H}_m^H \mathbf{H}_m + \sigma^2 \mathbf{I}_N \right)^{-1} \mathbf{y}_{m,n}'. \quad (20)$$

where $\sigma^2$ is the mean power of additive Gaussian noise which is known by the receiver, $\mathbf{I}_N$ is the $N \times N$ identity matrix, and $\mathbf{y}_{m,n}' \in \mathbb{C}^{K \times 1}$ has an element of $y_{m,n}'[k] = Y''[m,n,k]$.

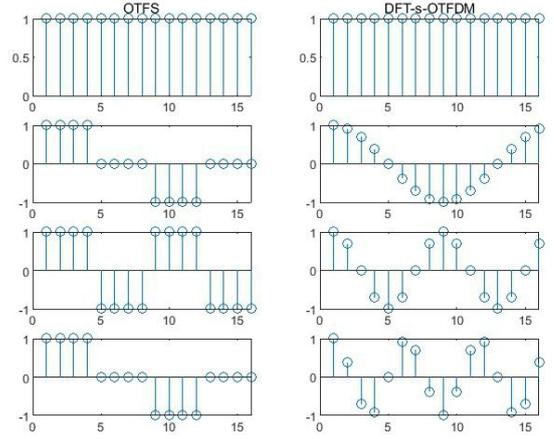

Fig. 2 The real parts of time-domain phase rotation coefficients in OTFS and DFT-s-OTFDM. $M = N = 4$.

The carried data at the receiver side is inevitably non-orthogonal in the doubly-selective channel. OTFDM keeps the orthogonality in the $M$-point Doppler replica, and deals with the non-orthogonality across Doppler replicas. The non-orthogonality can either be transformed into the orthogonality via time-domain equalization or solved by LMMSE. One interesting observation of the two Doppler despreading methods is that time-domain equalization is similar to frequency-domain equalization, while time-domain LMMSE is similar to spatial-domain LMMSE.

### C. OTFDM vs OTFS

If DFT-s-OTFDM employs an $M$-dimension DFT despreading, it will be very similar to OTFS. The difference is the Doppler dot product of Equation (8). With this step, the Doppler-dimension operations become more continuous than those of OTFS, which is also more similar to the practical Doppler effect. This feature makes the OTFDM transmitter equivalent to OFDM with comb interleaving. In another word, OTFDM utilizes the Doppler spreading feature already existed in OFDM. Moreover, the DFT-s-OTFDM sub-symbols have a perfect CP after the dot division, which ensures the sub-carrier orthogonality within the sub-symbol. The inter-sub-symbol interference is suppressed by Doppler despreading or other combining methods.

Another difference is that OTFDM employs a totally different receiver design, as OTFDM deals with the 2D dot-product channel in the TF domain, while OTFS deals with the 2D convolutional channel in the DD domain. Although DD domain shifting is closer to the nature of the practical channel, the TF domain dot product allows much more efficient processing. Similarly, the MIMO receiver uses a dot-product spatial channel model instead of an angle-domain pulse shift. For OTFDM MIMO, the 2D dot-product channel model can also be easily extended to the 3D dot-product channel model.

### IV. CHANNEL ESTIMATION AND PILOT OVERHEAD

#### A. Channel Estimation

The channel estimation is crucial for OTFDM as it requires 2D TF channel information for the demodulation. Although OTFDM can be equivalent to an $MN$-point OFDM with comb interleaving, OFDM is affected by the Doppler frequency and only provides the 1D frequency-domain channel estimation. To obtain TF 2D channel estimation, the processing of OTFDM sub-symbols is essential. The challenge is that the sub-symbol is not separable before Doppler despreading which has already relied on channel estimation. Therefore, OTFDM requires Doppler-domain guardbands to obtain the 2D TF channel information.

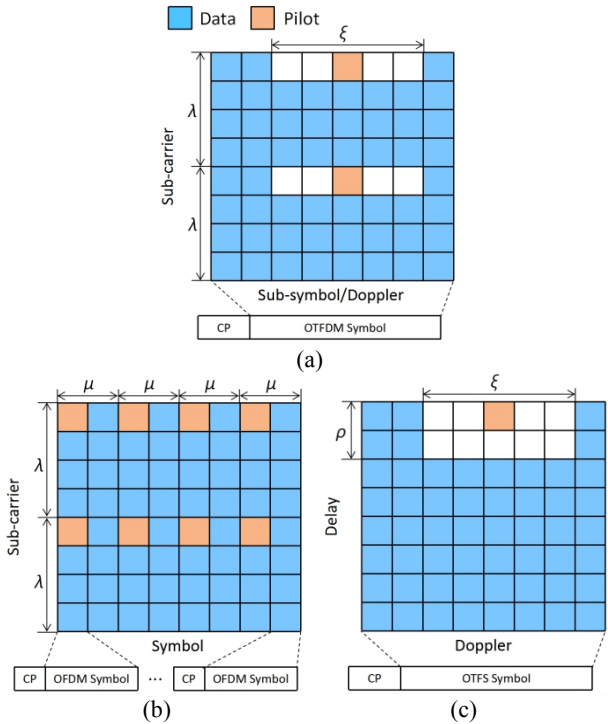

Fig. 3 The pilot insertion schemes of (a) OTFDM, (b) OFDM, and (c) OTFS.

The OTFDM pilot insertion is shown in Fig. 3(a). In the frequency domain, pilots are inserted with a density of $1/\lambda$, which also represents a repetition period of $\lambda$ sub-carriers. Similar to OFDM, $\lambda$ is decided by the severeness of frequency-selectivity. In the Doppler domain, the pilot is inserted with a width of $\xi$, the width is decided by the Doppler spread range, which reflects the severeness of time selectivity. In the practical off-grid model, $W$ also requires to be long enough to cover the sidelobe leakage.

In channel estimation, the OTFDM receiver first extracts the received pilot signal in the frequency and Doppler domain. At each pilot sub-carrier, the Doppler domain pilot is transformed into the time domain via an $N$-point IFFT of Doppler despreading. In this way, OTFDM obtains the coefficients of different Doppler replicas and pilot sub-carriers. These channel coefficients reflect the 2D TF selectivity. The OTFDM measures the dot-product frequency selectivity at the pilot sub-carrier and interpolates the frequency domain like what OFDM does. Moreover, the interpolation is also done among different OTFDM sub-symbols to obtain complete frequency-domain channel information of $\eta_{m,n,k}$ in Equation (16), as different sub-symbols locate at different frequencies as shown in Fig. 2.

### B. Pilot Overhead Analysis

Fig. 3(b) and (c) show the pilot insertion schemes of OFDM and OTFS. The pilot insertion of OTFDM is a fusion of that of OFDM and OTFS. Among all three schemes, the pilot insertion methods in both time domain and frequency domain can be divided into two kinds. One is to use a pulse, or Dirac delta function, to measure the shifting. The other is to use periodic sampling to measure the doubly-selectivity coefficients of the TF domain. Although the impulse response looks very intuitive for the impact of time and Doppler shift, it does not mean this is the best way of practical implementations.

For delay-domain pulse, the focus of power on one pulse in every $M$ points to acquire the channel causes severe PAPR issues. As a comparison, the frequency-domain periodic sampling spreads the pilot power over all time length. With one $M/\lambda$-point IFFT, the frequency-domain channel can be transformed into the time domain, which is in a $M/\lambda$-point overhead in the delay domain. The overheads using different methods, $M/\lambda$-point and $\rho$, are both required to be larger than $\max(\tau_i)/T_S$. Similarly, the Doppler domain pulse and time-domain period sampling have overheads of $\xi$ and $M/\mu$, both of which are required to be larger than two-fold of $\max(|v_i|)/\Delta f'$. To sum up, the pilot overheads are near, as the delay-domain and frequency-domain channel estimation can be easily transformed to each other via FFT and IFFT, and so does the Doppler-domain and time-domain.

## V. NUMERICAL RESULTS

### A. Simulation Settings

This simulation assumes a doubly-selective channel as described in Section II. The simulation parameters can be found in TABLE I. The channel model is based on TDL-C model in 5G NR. As TDL-C only describes the power and delay profile, a random Doppler shift is added for each path using a uniform distribution. For a fair comparison, OFDM and OTFDM use the common parameters including symbol length, CP length, and coding scheme. Also, the pilot overheads are set to be the same, which is inserted periodically in the frequency domain of the OFDM symbol and OTFDM sub-symbol. Note that at each OTFDM pilot sub-carrier, only one of the $W$ sub-symbols carries the pilot. The pilot power is set to be the power sum of $W$ pilots in each OFDM pilot sub-carrier to make the total pilot power equal and fair. In this paper, the interpolation is done via simple FFT-based oversampling. The CP-OFDM and CP-OTFDM are compared using low-complexity dot-division based equalization, which is very friendly to practical use.

TABLE I. LINK-LEVEL SIMULATION PARAMETERS

| Simulation Parameters | Value |
|---|---|
| Carrier frequency | 24 GHz |
| OFDM Sub-carrier spacing (SCS) $\Delta f$ | 15 kHz |
| OFDM Sub-carrier Number $MN$ | 4096 |
| OTFDM SCS $\Delta f'$ | 120 kHz |
| OTFDM Sub-carrier Number $M$ | 512 |
| OTFDM Sub-symbol Number $N$ | 8 |
| Pilot Overhead Density | 1/4 |
| OTFDM Pilot Width $W$ | 8 |
| CP length $N_{CP}$ | 288 |
| Modulation | QPSK |
| Channel Coding | LDPC |
| LDPC Code Rate | 1/2 |
| LDPC Block Size | 1024 |
| Signal Power to Noise Ratio (SNR) | 6 dB |
| Pilot Power Boosting | 6 dB |
| Channel Model | 5G TDL-C |
| Path Number $P$ | 24 |
| Root Mean Square Delay Spread | 1000 ns |
| Maximum Mobility | 500 km/h |

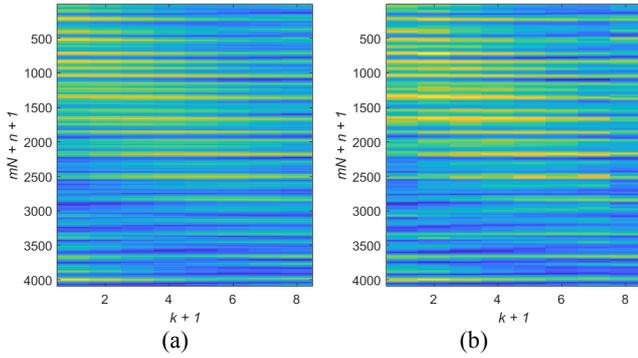

Fig. 4 (a) The ground-truth value and (b) interpolated channel estimation of 2D dot-product channel $H_m[n,k]$.

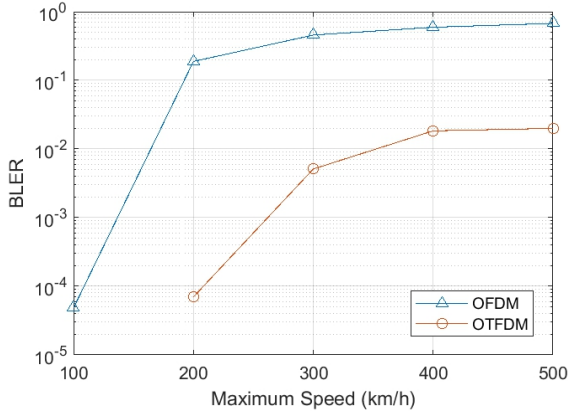

Fig. 5 Link-level performance comparison in doubly-selective channel with different maximum speeds.

### B. Results and Analysis

First, we take a look at the channel estimation of OTFDM. As mentioned before, OTFDM models a 2D dot-product channel $H_n[m,k]$ in Equation (14). Fig. 4 merges the 2D channel of all $N$ sub-symbols into one matrix and the mapping relationship can be found in the labels in the figure. For the ground-true value, the time-selectivity is calculated for each sub-symbol and sub-carrier. The channel estimation is obtained from the OTFDM pilot, which is interpolated to obtain the complete 2D dot-product channel as described before. As shown in Fig. 4, the interpolated channel estimation can well reflect the doubly-selective feature of the ground-true channel, which is very useful for the frequency- and time-domain equalization.

This simulation adopts simple time-domain equalization to deal with time-selectivity for OTFDM, which is consistent to frequency-domain equalization coping with frequency-selectivity. In Fig. 5, $2\times10^5$ drops are simulated to compare the block error rate (BLER) performance of OFDM and OTFDM. Different maximum speeds are used, and the random speeds of 24 paths follow a uniform distribution from 0 to the maximum. The BLER comparison shows OTFDM outperforms OFDM in the doubly-selective channel without sacrificing CP length. Taking BLER = $10^{-1}$ as a tolerable limit, the BLER of OFDM rapidly fails when the maximum speed is no less than 200km/h, while the BLER of OTFDM degrades slowly with the speed rising. OTFDM gains a BLER of around $2\times10^{-2}$ at a maximum speed of 500 km/h. The cost of OTFDM is around $N$-fold complexity as the dot-division equalization is now done in two dimensions.

### APPENDIX: THE PROOF OF OFDM EQUIVALENCE

In this appendix, the data part equivalence of OTFDM and comb interleaved OFDM is proved. With Equations (7) and (8), Equation (9) becomes Equation (A.1) at the bottom of this page. As

$$e^{j2\pi \frac{kmMN}{MN}} = 1,$$

Equation (A.1) becomes

$$S_3[l,k] = \frac{1}{\sqrt{MN}} \sum_{n=0}^{N-1} \sum_{m=0}^{M-1} e^{j2\pi \frac{nl+nkM+mlN+kmMN}{MN}} X[m,n]$$

$$= \frac{1}{\sqrt{MN}} \sum_{n=0}^{N-1} \sum_{m=0}^{M-1} e^{j2\pi \frac{(kM+l)(mN+n)}{MN}} X[m,n]. \quad (A.2)$$

Let $l' = kM+l$, $n' = mN+n$, Equation (A.2) becomes

$$s_3[l'] = \frac{1}{\sqrt{MN}} \sum_{n'=0}^{MN-1} e^{j2\pi \frac{n'l'}{MN}} x_T[n']. \quad (A.2)$$

Where $x_T[n']$ is the $n'$-th element of $\text{vec}(\mathbf{X}^T)$. This equivalence shows that the data part in $N$ sub-symbols are interleaved using a comb shape with different sub-carrier frequency shift $0, \Delta f, ..., (N-1)\Delta f$.

$$S_3[l,k] = \frac{1}{\sqrt{N}} \sum_{n=0}^{N-1} S_2[l,n] e^{j2\pi \frac{nk}{N}} = \frac{1}{\sqrt{N}} e^{j2\pi \frac{nl}{MN}} \sum_{n=0}^{N-1} S_1[l,n] e^{j2\pi \frac{nk}{N}} = \frac{1}{\sqrt{N}} e^{j2\pi \frac{nl}{MN}} \sum_{n=0}^{N-1} \left( e^{j2\pi \frac{nk}{N}} \frac{1}{\sqrt{M}} \sum_{m=0}^{M-1} X[m,n] e^{j2\pi \frac{ml}{M}} \right)$$

$$= \frac{1}{\sqrt{MN}} \sum_{n=0}^{N-1} \sum_{m=0}^{M-1} e^{j2\pi \left(\frac{nl}{MN}+\frac{nk}{N}+\frac{ml}{M}\right)} X[m,n] = \frac{1}{\sqrt{MN}} \sum_{n=0}^{N-1} \sum_{m=0}^{M-1} e^{j2\pi \frac{nl+nkM+mlN}{MN}} X[m,n]. \quad (A.1)$$